\documentclass[12pt]{iopart}

\usepackage{iopams}
\usepackage{color}
\expandafter\let\csname equation*\endcsname\relax
\expandafter\let\csname endequation*\endcsname\relax
\usepackage{amsmath}
\usepackage{graphicx}
\begin{document}
\title{Effects of excitation frequency on high-order terahertz sideband generation in semiconductors}

\author{Xiao-Tao Xie}
\address{Department of Physics, The Chinese University of Hong Kong, Shatin, New Territories, Hong Kong, China}
\address{Department of Physics, Northwest University, Xi'an 710069, China}
\author{Bang-Fen Zhu}
\address{Department of Physics, Tsinghua University, Beijing 100084, China}
\author{Ren-Bao Liu$^*$}
\address{$^*$ Corresponding author (email: {\tt rbliu@phy.cuhk.edu.hk})}
\address{Department of Physics, The Chinese University of Hong Kong, Shatin, New Territories, Hong Kong, China}

\begin{abstract}
We theoretically investigate the effects of the excitation frequency on the plateau of high-order terahertz sideband
generation (HSG) in semiconductors driven by intense terahertz (THz) fields.
We find that the plateau of the sideband spectrum strongly depends on the detuning between the NIR laser field and the band gap. We use the quantum trajectory theory (three-step model) to understand the HSG. In the three-step model, an electron-hole pair is first excited by a weak laser, then driven by the strong THz field, and finally recombine to emit a photon with energy gain. When the laser is tuned below the band gap (negative detuning), the electron-hole generation is a virtual process that requires quantum tunneling to occur. When the energy gained by the electron-hole pair from the THz field is less than 3.2 times the ponderomotive energy, the electron and the hole can be driven to the same position and recombine without quantum tunneling, so the HSG will have large probability amplitude. This leads to a plateau feature of the HSG spectrum with a high-frequency cutoff at about 3.2 times the ponderomotive energy above the band gap. Such a plateau feature is similar to the case of high-order harmonics generation in atoms where electrons have to overcome the binding energy to escape the atomic core. A particularly interesting excitation condition in HSG is that the laser can be tuned above the band gap (positive detuning), corresponding to the unphysical ``negative" binding energy in atoms for high-order harmonic generation. Now the electron-hole pair is generation by real excitation, but the recombination process can be real or virtual depending on the energy gained from the THz field, which determines the plateau feature in HSG. Both the numerical calculation and the quantum trajectory analysis reveal that for positive detuning, the HSG plateau cutoff depends on the frequency of the excitation laser. In particular, when the laser is tuned more than 3.2 times the ponderomotive energy above the band gap, the HSG spectrum presents no plateau feature but instead sharp peaks near the band edge and near the excitation frequency.
\end{abstract}
\maketitle

\section{Introduction}

High-order harmonic generation (HHG) results from the interaction of an intense laser with atoms
or molecules. HHG provides a mechanism for generating coherent extreme ultraviolet (XUV) and X-ray
attosecond pulses \cite{Brabec,Krause1}.
The three-step model was established to describe the physical processes of HHG \cite{Krause2,Corkum,Lewenstein}:
The strong laser field tilts the binding potential and the electron escapes from the
charged core of the atom or molecule through quantum tunneling; the electron
is then accelerated in the free space by the laser field; when the electron recollides with the charged core,
a very energetic photon is emitted. Recently high-order teraherz sideband generation (HSG)
in semiconductors was predicted \cite{Liu2}, which has a physical mechanism similar to HHG but occurs
at a very different frequency range. In semiconductors, an electron can be excited from
the valance band to the conduction band with a hole left behind. The recollisions between energetic
holes and electrons accelerated by a strong THz field result in HSG. Recently, HSG
has been experimentally demonstrated \cite{Liu,Zaks}. The excitonic effect has also
been theoretically studied \cite{Yan}.

A fundamental difference between HHG in atoms and HSG in semiconductors is that the
electron-hole (e-h) pairs in HSG are elementary excitations caused by NIR lasers. In HHG,
the electrons need to overcome the binding energy by quantum tunneling and therefore the
quantum trajectories that satisfy the least action condition have only complex solutions.
In HSG, the laser frequency $\Omega$ can be tuned from below to above the semiconductor
bandedge $E_g$, and correspondingly, the initial energy of the  e-h pairs generated
by the NIR laser can be tuned from negative to positive relative to the bandedge. Particularly,
for e-h pairs with positive excess energy ($\Omega-E_g>0$, corresponding to ``negative''
binding energy of atoms in HHG), the e-h pairs can be directly created by
real excitation without quantum tunneling, and in turn the quantum trajectories may
have real solutions. On the other hand,  the laser can be tuned so high above the band
edge that  the initial velocities of the e-h pairs are too high for the THz field
to drive electrons and holes into recollision. Then recombination of e-h pairs  has to
occur through quantum tunneling (c.f. quantum tunneling in creation of e-h pairs for
laser tuned below band edge). Therefore, there will also be a cutoff of HSG on the
excitation laser frequency.

Such consideration motivates us to investigate the dependence of the HSG in semiconductors,
in particular, its plateau features, on the frequency of the NIR laser. We calculate the HSG
for various excitation laser frequencies and examine the plateau features. The HSG spectrum
is explained by quantum trajectories in the three-step model.
The previous studies on HSG for negative detuning ($\Omega-E_g<0$) show that the maximal
and minimal orders of the sideband plateau are given by the cut-off law \cite{Liu2}
\begin{subequations}
\begin{eqnarray}
E_{max}&=&(E_g-\Omega+3.17U_p)/\omega_0~, \label{e0a}\\
E_{min}&=&0~, \label{e0b}
\end{eqnarray}
\end{subequations}
where $U_p=F^2/4\omega_0^2$ is the ponderomotive energy, with $F$ and $\omega_0$ being
the strength and frequency of the THz field, respectively. The laser detuning $E_g-\Omega$ plays a role
similar to the binding energy in atoms for HHG \cite{Liu2}.
The cut-off frequencies can be derived from a classical calculation with the assumption that
the electrons tunnel to the conduction band with zero initial velocity. In this paper, we will
show that for positive laser detuning ($\Omega-E_g>0$) the HSG plateau cutoffs do not satisfy the Eqs.~(\ref{e0a})
and (\ref{e0b}) and depend on the detuning. This characteristic indicates that the HSG
spectrum may be modified by tuning the frequency of the excitation laser.

\section{Model and Numerical Simulation }

Under an intense THz field, the kinetic energy acquired by the e-h pair can be much greater than the exciton binding energy in the semiconductor and the amplitude of the relative motion can be much greater than the exciton radius, so the essential physics of the HSG can be grasped by the motion of free electrons and holes without Coulomb interaction \cite{Liu,Yan}. The equation of relative motion for the electron-hole pair without coulomb interaction is \cite{Liu2}
\begin{eqnarray}\label{e1}
i\frac{\partial }{\partial t} \Psi (\mathbf{r},t) = [(\mathbf{p}-\mathbf{A})^2 + E_g ]\Psi (\mathbf{r},t) + \mathbf{d}\cdot \mathbf{E}(t) \delta(\mathbf{r}),
\end{eqnarray}
where $\mathbf{p}$ and $\mathbf{d}$ are the momentum and the interband dipole matrix element, respectively. $\mathbf{A}(t)=-\frac{F}{\omega_0}\sin(\omega_0 t)\hat{e}_z$
is the vector potential of the intense THz field, and $\mathbf{E}(t) = \mathbf{E}_{NIR}\exp(-i\Omega t)$
is the NIR laser field. Here we have assumed that the electron and the hole are generated at
the same position ($\mathbf{r}=0$) and the dipole matrix element $\mathbf{d}$ is a constant independent of the momentum.
The optical polarization of the e-h pair is $\mathbf{P}(t) = -\mathbf{d}^*\psi(0,t)$.
The interband polarization can be expressed as
\begin{eqnarray}\label{e2}
\mathbf{P}(t) = i\int \mathbf{d}^* \delta(\mathbf{r}) K(\mathbf{r}t,\mathbf{r}'t') \theta(t-t'){\mathbf d}\cdot{\mathbf E}(t')\delta(\mathbf{r}')d\mathbf{r} d\mathbf{r}' dt' ~,
\end{eqnarray}
where $\theta(x)$ is the Heavise step function, and the propagator
\begin{eqnarray}\label{e3}
&& K(\mathbf{r}t,\mathbf{r}'t')=\langle \mathbf{r}| \mathbf{k} \rangle \langle \mathbf{k} | K (\mathbf{k}t,\mathbf{k}'t')| \mathbf{k}' \rangle\langle \mathbf{k}' |  \mathbf{r}' \rangle\nonumber\\
&&~~~~=(2\pi)^{-3} \int d\mathbf{p}\exp\left[ i\mathbf{p}\cdot(\mathbf{r}-\mathbf{r}')-iE_g(t-t')-i\int_{t'}^{t}dt''[\mathbf{p}-\mathbf{A}(t'')]^2  \right]~.\nonumber\\
\end{eqnarray}
Using the Fourier transform,
we get the polarization strength of the $N$th order sideband
\begin{eqnarray}\label{e4}
P_{N} = i \mathbf{d}^*\mathbf{d} \cdot \int \frac{dt d\tau d\mathbf{p}}{(2\pi)^{3}} \exp\left[  iS  \right] \theta(\tau) {\mathbf E}(t-\tau)~,
\end{eqnarray}
where the action is
\begin{eqnarray}\label{e5}
S(\mathbf{p},t,\tau) = -\int_{t-\tau}^{t}dt''[\mathbf{p}-\mathbf{A}(t'')]^2 + (\Omega-E_g)\tau + N\omega_0 t~.
\end{eqnarray}
Here $\tau$ denotes the delay between the recombination and the creation of the e-h pair.
The action $S(\mathbf{p},t,\tau)$ is a phase introduced by the motion of the  e-h  pair
with canonical momentum $\mathbf{p}$. Due to the inversion symmetry, only the even-order sidebands
(with even $N$) can be generated~\cite{Shen}.

For a strong THz field, the e-h pair performs relative motion with amplitude
much greater than its wavepacket diffusion range. Therefore the motion is well grasped
by a few quantum trajectories that satisfy the stationary phase conditions. The
stationary-phase points are determined by a set of saddle-point equations
\begin{subequations}
\begin{eqnarray}
\partial_p S = 0 &\Longrightarrow&  \mathbf{p}\tau - \int_{t-\tau}^{t} \mathbf{A}(t')dt' =0~,\label{e6}\\
\partial_\tau S = 0 &\Longrightarrow& [\mathbf{p}-\mathbf{A}(t-\tau)]^2 = \Omega -E_g~,\label{e7} \\
\partial_t S = 0 &\Longrightarrow& [\mathbf{p}-\mathbf{A}(t)]^2 - [\mathbf{p}-\mathbf{A}(t-\tau)]^2 = N\omega_0~.\label{e8}
\end{eqnarray}
\end{subequations}
The physical meanings of these equations are: Eq.~(\ref{e6}) -- return of the electron to the position of the
hole after acceleration by the THz field [the velocity integrated over time
equals zero]; Eq.~(\ref{e7}) -- energy conservation at the excitation of the e-h pair; Eq.~(\ref{e7}) --
energy conservation for the sideband generation through recombination of the e-h pair.

The laser detuning $\Omega -E_g$ determines the mechanism of the e-h generation and recombination
(with or without quantum tunneling). If $\Omega < E_g$, the electron enters into the continuum
via quantum tunneling assisted by the strong THz field. The tunneling physics results in complex
solutions of $t$ and $\tau$, which in turn leads to reduced HSG intensity. When $\Omega \geq E_g$,
real excitation of e-h pairs is possible and so are real solutions of $t$ and $\tau$. Without
requiring quantum tunneling through a binding energy barrier, the HSG can have large amplitudes.
However, if the initial energy and hence the initial relative velocity of the e-h pair are too
large ($\Omega-E_g>3.17U_p$), the THz field would not be able to drive the electron and the hole back
to the same position. Then only through quantum tunneling can the e-h pair recombine,
which means the saddle point equations would not have real solutions. So the HSG amplitude will
be large when the excitation energy $0<\Omega-E_g < 3.17U_p$ and drops rapidly outside this range
-- the HSG presents a plateau feature not only in the emission spectrum but also in the excitation
spectrum. Moreover, as we will show below, the HSG emission plateau depends on the laser
excitation frequency.

Substituting Eq.~(\ref{e6}) into Eqs.~(\ref{e7}) and ~(\ref{e8}), we get the saddle point equations
with respect to $t$ and $\tau$
\begin{subequations}\label{e90}
\begin{eqnarray}
&& \frac{\Omega-E_g}{4U_p} =  \left[\alpha\sin(\omega_0 t-\frac{\omega_0 \tau}{2}) -\beta\cos(\omega_0 t-\frac{\omega_0 \tau}{2})\right]^2~, \label{e9}\\
&& \frac{N\omega_0 + \Omega-E_g}{4U_p}  =  \left[\alpha\sin(\omega_0 t-\frac{\omega_0 \tau}{2})  + \beta\cos(\omega_0 t-\frac{\omega_0 \tau}{2})\right]^2~, \label{e10}
\end{eqnarray}
\end{subequations}
where
\begin{eqnarray}
\alpha(\tau) = \cos\frac{\omega_0\tau}{2}-\frac{\sin\frac{\omega_0\tau}{2}}{\omega_0\tau/2},~~~\beta(\tau)=\sin\frac{\omega_0\tau}{2}.\nonumber
\end{eqnarray}
The equations can have real solutions if $\Omega > E_g$ and not too much above the band edge,
which is not possible for HHG in atoms where the binding energy has to be positive \cite{Becker}.

\begin{figure}
\includegraphics[width=\textwidth]{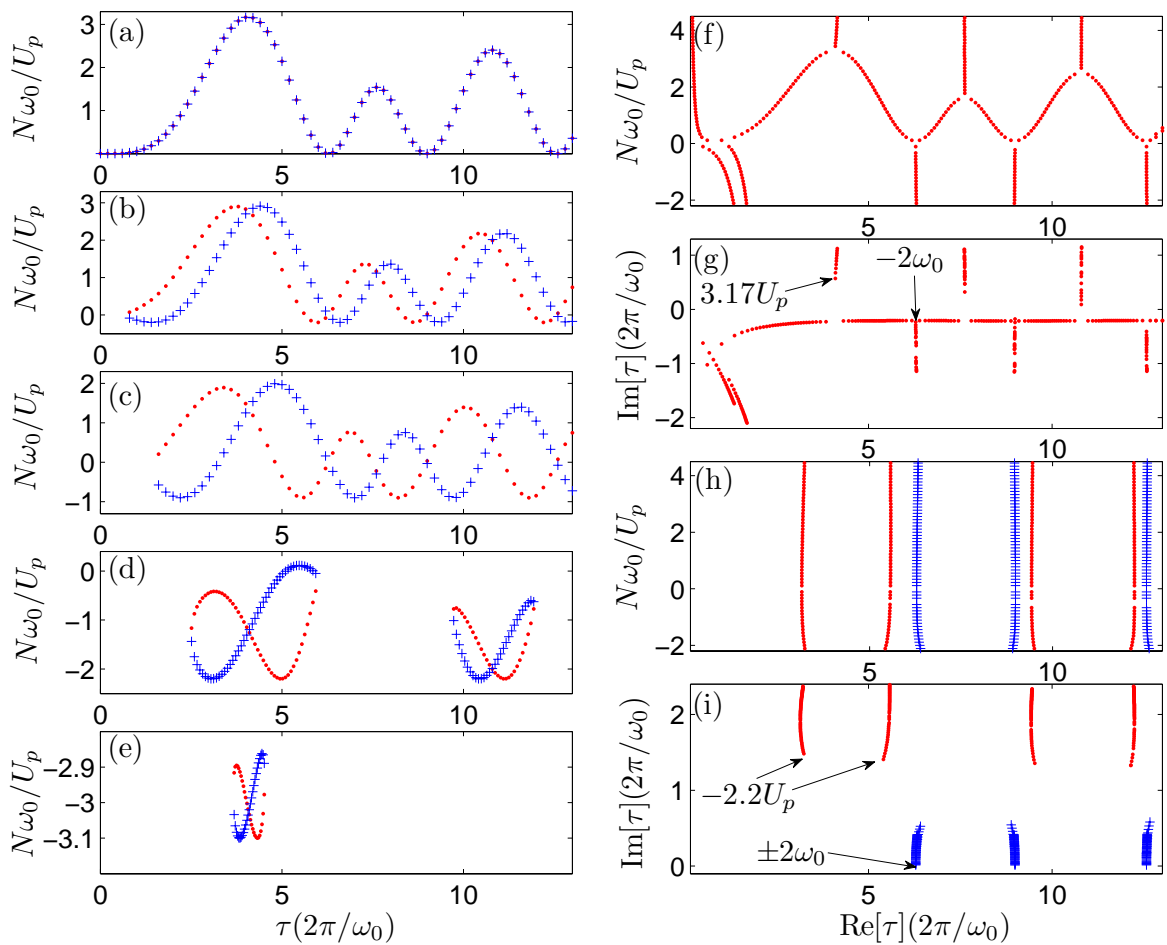}
\caption{ The dimensionless sideband frequency shift $N\omega_0/U_p$ in the plateau region as a function of the
return time $\tau$ for various NIR laser detuning $\Delta=\Omega-E_g$. (a) $\Delta=0.0$; (b) $\Delta=0.2U_p$;
(c) $\Delta=0.9U_p$; (d) $\Delta=2.2U_p$; (e) $\Delta=3.1U_p$; (f) $\Delta=-0.1U_p$; (g) $\Delta=-0.1U_p$;
(h) $\Delta=4.4U_p$; (i) $\Delta=4.4U_p$. The dots and pluses in (a)-(e) are for initial relative velocity of
the e-h pair anti-parallel and parallel to the THz field ${\mathbf E}(t-\tau)$, respectively. In (f)-(i), the
saddle point solutions of $\tau$ are complex numbers and the results are plotted as functions of the real part
of $\tau$. The photon energy and the strength of the THz field are $5~\mathrm{meV}$ and $30~\mathrm{KV/cm}$,
respectively. The effective mass of the e-h pair is set as $0.076~\mathrm{m_e}$. These parameters give
$U_p=90~\mathrm{meV}$. ~\label{f1}}
\end{figure}

\begin{figure}
\includegraphics[width=\textwidth]{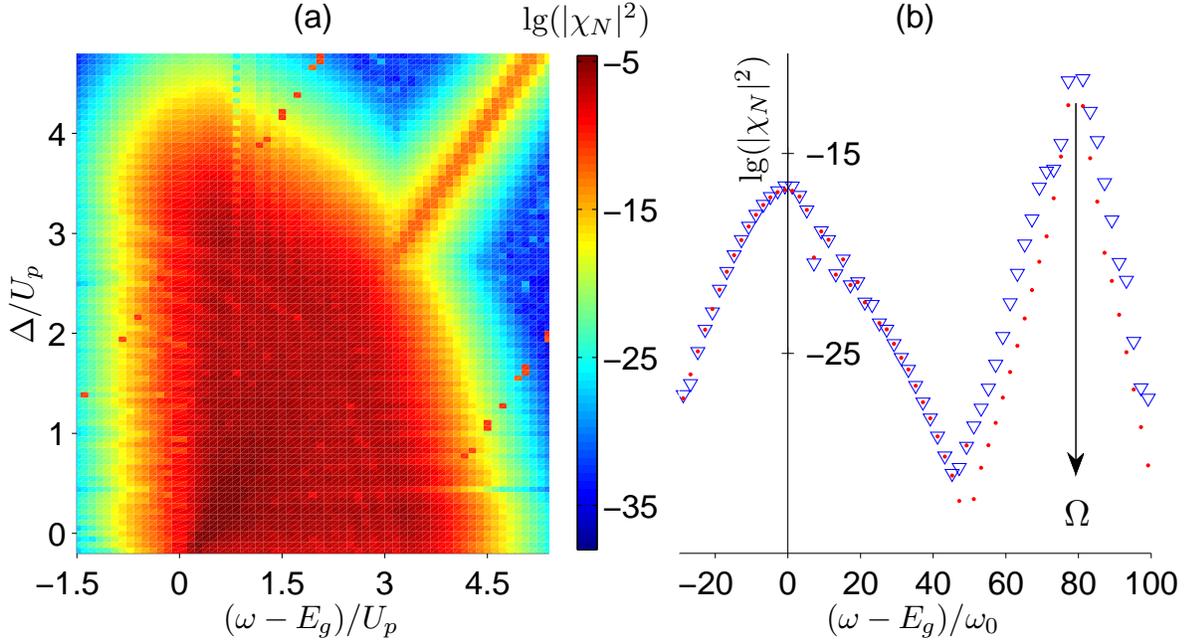}
\caption{HSG intensity as a function of excitation and emission frequencies. (a) The contour plot of the
sideband strength $|\chi_N|^2$ as a function of detuning $\Delta=\Omega-E_g$ and sideband frequency $\omega$.
Here the sideband strength is exactly calculated. The width of the sideband plateau depends on $\Delta$.
(b) The sideband strength $|\chi_N|^2$ calculated by the saddle-point method (triangles) and exactly (dots)
with detuning above the $3.17U_p$ threshold ($\Delta=400~\mathrm{meV}\simeq 4.4U_p$). The parameters are
the same as in Fig.~\ref{f1}~\label{f2}}
\end{figure}

Equation (\ref{e90}) provides some insight to understand the HSG spectrum. The e-h pair gains or loses
kinetic energy to generate sidebands. The dimensionless sideband shift frequency $N\omega_0/U_p$
in the plateau regime is shown in Fig.~\ref{f1} as a function of the delay time for various NIR
laser detuning. In our calculation, the effective mass of the e-h pair is chosen to be $0.076~\mathrm{m_e}$
as in GaAs, the photon energy of the THz field is $5~\mathrm{meV}$ and its strength is $30~\mathrm{KV/cm}$.
Such parameters above yield $U_p=90~\mathrm{meV}$. As shown in Fig.~\ref{f1}, the upper and lower
cutoff orders of the HSG plateau depend on the detuning $\Delta \equiv \Omega-E_g$.
 The saddle
point $t_n$ and $\tau_n$ are real if the laser detuning $\Delta$ is such a range that the e-h pair has trajectories
satisfying the the classical mechanics. If the laser detuning is outside this range,
$t_n$ and $\tau_n$ are complex numbers and the trajectories have to be assisted by quantum tunneling.
For positive laser detuning ($\Delta>0$), the initial relative velocity of the e-h pair is non-zero and can be
along different directions. To fulfill the saddle point equation (\ref{e6}) (the condition that the
electron and the hole return to the same position), the initial velocity has to be anti-parallel or
parallel to the THz electric field ${\mathbf E}(t)$. Therefore for $\Delta>0$ the kinetic energy the
e-h pair can acquire from the THz field is a double-valued function as shown in Fig.~\ref{f1}.
As shown as gaps in Figs.~\ref{f1}~(d)-(e), the real saddle point equations have no real
solutions for certain ranges of delay time $\tau$ even for laser detuning $\Delta$
$\in [0,~3.17U_p]$. This is because when the initial velocity of the e-h
pair is large, the THz field cannot drive the electron and the hole back into
the same position unless the delay time is in the proper ranges.
For negative detuning ($\Delta<0$), the e-h pair is generated by quantum tunneling.
The saddle-point solutions of $\tau$ are in general
complex numbers [see Fig.~\ref{f1} (f,g)].
When the emission frequency outside the plateau region ($N\omega_0>3.17Up$ or $N<0$)
goes away from the cut-off frequencies, the imaginary part of the saddle point $\tau_n$ is large
and increases rapidly [see Fig.~\ref{f1} (g)]. That means small quantum tunneling rate and
hence weak sidebands. When $\Delta \geq 3.17U_p$.
When the laser detuning is positive and $\Delta \geq 3.17U_p$,
the initial velocity of the e-h generated by real excitation is so large that the THz field cannot
bring the electron and the hole to the same position again for recombination. The saddle-point solutions are also complex.
The corresponding saddle point solutions of $\tau$ are depicted in Fig.~\ref{f1} (h) and (i).
When the emission frequency goes away from the excitation frequency (blue star symbols) or the band edge (red dot symbols),
the imaginary parts of the saddle points of $\tau$ [Fig.~\ref{f1} (i)] increases rapidly, so the sideband strength $P_N$
decreases rapidly and present sharp peaks at the band edge and the excitation frequency (see Fig.~\ref{f2} (b)).


We now show the quantum trajectory approach to calculating the HSG spectrum. By solving the
saddle-point equations above, we obtain the saddle points ($t_n$, $\tau_n$). The action is then
\begin{eqnarray}
S_{cl}(t,\tau)= N\omega_0 t + (\Omega-E_g-2U_p)\tau + 2U_p\tau\gamma^2 + 2U_p\tau\alpha\gamma\cos(2\omega_0 t - \omega_0\tau) ~,\nonumber\\
\label{e12}
\end{eqnarray}
where $\gamma = \sin(\omega_0\tau/2)/(\omega_0\tau/2)$. After integrating the polarization strength $P_{N}$ with Gaussian integral, the susceptibility is determined
by \cite{Yan}
\begin{eqnarray}
\chi_{N} = \frac{\omega P_{N}}{\pi E_{NIR}} \approx \sum_n \frac{i \mathbf{d}^*\mathbf{d}}{(\sqrt{4\pi\tau_n i + 0^+})^3}
 \exp\left[ i S_{cl}(t_n,\tau_n)  \right] \sqrt{\frac{4\pi^2i^2}{\det S_{cl}''}}~.\label{e13}
\end{eqnarray}
Here $S''$ represents the second-order derivative Jacobian determinant of the corresponding action.
The results calculated by exact solution and the saddle-point method are shown in Fig.~\ref{f2}.
The characteristic of the HSG spectrum agrees well with the results in Fig.~\ref{f1}: the
plateau shrinks when the detuning is increased. The saddle points $\tau_n$ and
$t_n$ are complex outside the plateau region, which indicates that the quantum tunneling occurs
in the emission. The lower cut-off frequency of the HSG plateau is the band edge of the semiconductor
since the final kinetic energy of the e-h pair has to be positive. As discussed above, when the laser
detuning is above the $3.17U_p$ threshold, the plateau feature vanishes and instead the sideband
intensity presents two rapidly dropping peaks at the laser frequency and the band edge.

\begin{figure}
\includegraphics[width=\textwidth]{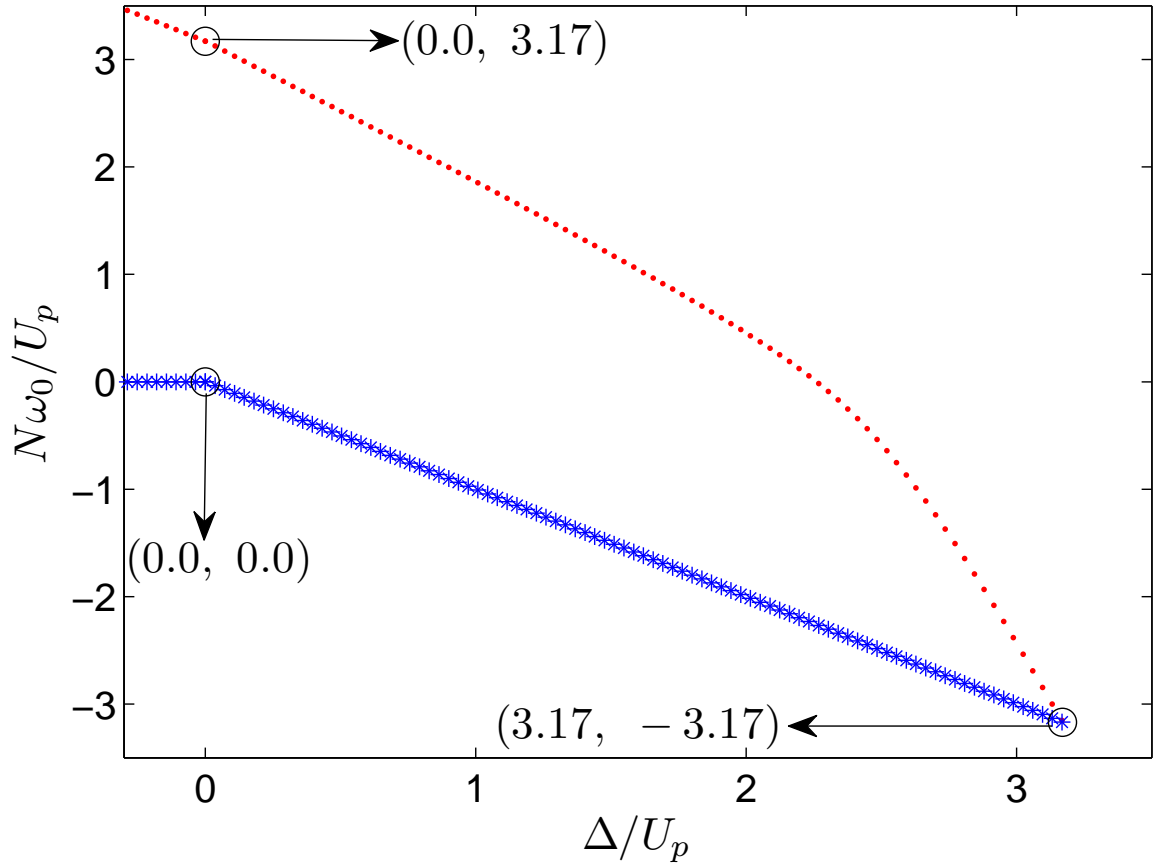}
\caption{ The dependence of the cut-off frequency of the sideband plateau on $\Delta/U_p$. The minimal and maximal orders of the cut-off frequencies are denoted by stars and dots, respectively. ~\label{f4}}
\end{figure}

Figure~\ref{f4} summarizes the width of the HSG sideband as a function of the laser detuning $\Delta$.
By searching the real roots of the saddle points $\tau_n$ and $t_n$ for Eqs.~(\ref{e9}) and (\ref{e10}),
we obtain the lower and upper cuttoffs of the HSG spectrum. As shown in Fig.~\ref{f4},
the width of the HSG plateau decreases with increasing of the laser detuning $\Delta$. But
the variation is rather slow in the range $\Delta/U_p \in (0,~2.2)$.

\section{Summary}

In summary, we have studied how hight-order THz sideband generation in semiconductors
driven by intense THz field depends on the detuning of the NIR laser from the semiconductor band edge.
As compared with HHG in atoms, the HSG can be studied for positive laser detuning (laser above
the semiconductor band edge), corresponding to unphysical ``negative'' binding energy for HHG in atoms.
Exact numerical simulation shows that for positive detuning, the HSG plateau
shrinks with increasing the laser detuning and eventually vanishes when the laser is higher than the $3.17U_p$
threshold above the semiconductor band edge. Such features are well understood using the quantum
trajectory approach.

\section{Acknowledgments}
This work is supported in part by Hong Kong RGC/GRF Project 401512 and the Hong Kong
Scholars Program (Grant No. XJ2011027). X.-T.X. would like to thank the National Natural
Science Foundation of China (Grant No. 61008016).



\end{document}